\documentstyle[aps,times,eqsecnum,epsf,12pt,amsmath,amssymb,amsbsy]{revtex}

\tighten
\begin{document}
\title{Phase Coexistence of Complex Fluids in Shear Flow} \author{\bf
  Peter~D. Olmsted$^{1}$ and C.-Y.~David Lu$^{2}$}
\address{$^1$Department of Physics, University of Leeds, Leeds LS2
  9JT, UK {\bf\tt (p.d.olmsted@leeds.ac.uk)}
  and $^2$Departments of Physics and Chemistry \& Center of Complex 
  Systems, \\
  National Central University, Chung-li, 320 Taiwan {\bf\tt
    (dlu@joule.phy.ncu.edu.tw)}} \date{\today} \maketitle
\def\thefootnote{\fnsymbol{footnote}}
\vskip10truept
\hrule
\vskip10truept
\begin{abstract} 
  We present some results of recent calculations of rigid rod-like
  particles in shear flow, based on the Doi model. This is an ideal
  model system for exhibiting the generic behavior of shear-thinning
  fluids (polymer solutions, wormlike micelles, surfactant solutions,
  liquid crystals) in shear flow. We present calculations of phase
  coexistence under shear among weakly-aligned (paranematic) and
  strongly-aligned phases, including alignment in the shear plane and
  in the vorticity direction (log-rolling). Phase coexistence is
  possible, in principle, under conditions of both common shear stress
  and common strain rate, corresponding to different orientations of
  the interface between phases. We discuss arguments for resolving
  this degeneracy. Calculation of phase coexistence relies on the
  presence of inhomogeneous terms in the dynamical equations of
  motion, which select the appropriate pair of coexisting states. We
  cast this condition in terms of an equivalent dynamical system, and
  explore some aspects of how this differs from equilibrium phase
  coexistence.
\end{abstract}
\vskip10truept
\hrule
\section{Introduction}
\begin{figure}
  \hsize\columnwidth\global\linewidth\columnwidth
  \displaywidth\columnwidth \epsfxsize=5.0truein
\centerline{\epsfbox{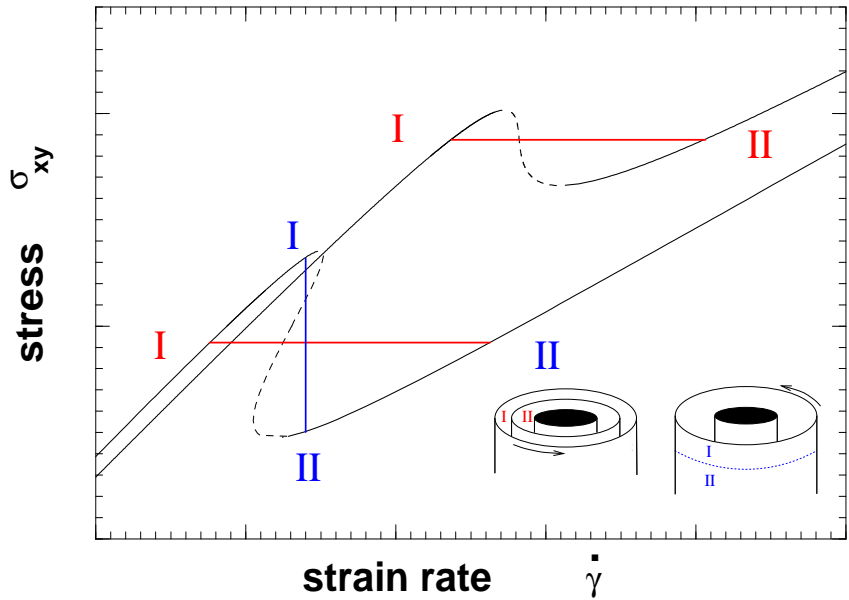}}
\caption{ Stress--strain-rate curves for the Doi
  model (from Fig.~\protect{\ref{fig:constit}}) for two
  concentrations. The dashed lines are unstable steady states.
  Straight lines indicate possible coexistence (at the same
  concentration) with either common stress (horizontal lines) or
  strain rate (vertical line) in the two phases.  Inset: geometries
  for common stress (left) or common strain rate (right) coexistence.}
\label{fig:both}
\end{figure}

Shear flow induces phase transitions and dynamic instabilities in many
complex fluids, including wormlike micelles
\cite{berret94a,berret94b,BPD97,Capp+97}, liquid crystals,
\cite{hess76,olmsted90,olmsted92,mather97,see90,olmstedlu97,safinya91};
and lamellar surfactant systems which can ``roll'' into multilamellar
vesicles (``onions'') \cite{catesmilner89,roux93,diat93,diat95}.
These instabilities typically manifest themselves in non-monotonic
constitutive curves such as those in Fig.~\ref{fig:both}
\cite{doiedwards,catesmcleish93,spenley93,cates90,rehage91,mcleish86},
and in several systems, including wormlike micelles and lamellar
surfactant solutions, are accompanied by observable coexistence of two
macroscopic ``phases'' of material.

If a mean strain rate forces the system to lie on an unstable part of
the flow curve (with negative slope), the system can phase separate
into regions (bands) with high and low strain rates and still maintain
the applied strain rate
\cite{olmsted92,spenley93,berret94a,BritCall97}.  Fig.~\ref{fig:both}
shows that phase separation can occur at either common stress or
common strain rate, depending on the geometry of the bands
\cite{schmitt95}: bands stacked along the axis of a Couette cell have
the same strain rate and different shear stresses, while radial phase
separation imposes a uniform shear stress and different strain rates.
The shear-thinning wormlike micelle system phase separates radially
into common-stress bands, while shear-thickening systems have been
observed to separate into bands with both the common strain rate
(worms) \cite{boltenhagen97a,boltenhagen97b} or common stress (worms
and onions) \cite{WFF98,Bonn+98} geometries, although the evidence for
true steady state phase separation at common stress is not yet firm.

Other systems with flow-induced ``phase transitions'' include
colloidal suspensions of plate-like particles (which shear thicken and
sometimes crystallize) \cite{brown98} or nearly monodisperse spheres
\cite{haw98}, as well as a variety of surfactant-like solutions of
diblock copolymers in selective solvents \cite{DPB96}. With so many
increasingly detailed and careful experiments on so many systems, it
would be nice to have a consistent framework for non-equilibrium
transitions. Unfortunately, most systems are sufficiently complicated
that none of the observed transitions can be completely described,
even qualitatively, by a credible microscopic model. For example, in
certain limits the class of shear-thinning wormlike micelle systems
which shear band has a mature theory for the linear rheology, which
may be extended using successful ideas from entangled polymer dynamics
to predict an instability \cite{spenley93}.  Unfortunately, a complete
description of phase coexistence also requires knowledge of the
shear-induced state, as well as details of the concentration
dependence and, as we shall see, the inhomogeneous contribution to the
dynamical equations.  We are, at present, far from having all of these
ingredients, and in many cases we do not have a clear understanding of
even the structure of the high shear rate state, much less its
dynamics.

Recently, we have studied a well-known model, the Doi model for rigid
rod suspensions in shear flow \cite{doi81}, which, while admittedly
the product of many approximations, provides physically well-founded
dynamics for both quiescent and shear-induced states. Although the
shear rates necessary for inducing a transition are, in practice,
typically quite high unless very long rods are used, and physical
systems are often susceptible to various dynamical instabilities, this
model system is quite helpful for building intuition about how to
calculate non-equilibrium ``phase diagrams'' and how their resulting
topologies resemble and differ from their equilibrium counterparts.

A vexing question for non-equlibrium calculations is how to replace
the free energy minimization familiar from equilibrium thermodynamics,
to determine the analog of a first order phase transition. In the
context of Fig.~\ref{fig:both}, one needs to determine the selected
stress for an imposed strain rate (for phase separation at a common
stress). It has emerged that an unambiguous resolution of this problem
is to include explicit non-local terms in the dynamical equations and
explicitly construct the coexisting state
\cite{kramer81a,olmsted92,spenley96,pearson94,olmstedlu97}.  This
reduces to the equilibrium construction in the case of zero shear, and
can be shown to yield a single (barring accidental degeneracies)
stress (given all other imposed conditions) at which coexistence
occurs \cite{jsplanar}.

Below we summarize some results of our calculations on the Doi model
for rigid rod suspensions in shear flow \cite{olmstedlu97}; the
details will be published elsewhere \cite{tobepub98}. This system is
surprisingly rich, given its apparent simplicity and fairly obvious
coupling of internal order to flow. Then we discuss some aspects of
the interface construction for determining coexistence, and how it
compares to its equilibrium counterpart.

\section{The Doi Model}
The modified Doi model \cite{doi81,olmstedlu97} describes the dynamics
of the rod-like particles suspension. The orientational
degrees of freedom are parametrized by the conventional liquid
crystalline order parameter tensor 
\begin{equation}
Q_{\alpha\beta}({\bf r}) = \langle \nu_{\alpha}\nu_{\beta} - 
\case13\delta_{\alpha\beta}\rangle,
\end{equation}
where $\langle\cdot\rangle$ denotes an average around the point ${\bf r}$ of 
the second moment of the rod orientations $\boldsymbol{\nu}$. For
rigid rods the phase diagram, and in fact the dynamics, can be more
conveniently represented by the exclude volume parameter $u$, defined
by
\begin{equation}
u=\phi L {\alpha},
\end{equation}
where $\phi$ is the rod volume fraction, $L$ is the rod aspect ratio and $\alpha$ is an ${\cal O\/}(1)$
prefactor \cite{doi81}. Beginning from the Smoluchowski equation for a
solution of rigid rods, and including a Maier-Sauper--like
orientational interaction parameter, Doi was able to derive
approximate coupled equations of motion for the dynamics of
$\boldsymbol{Q}$, and the fluid velocity ${\bf v}({\bf r})$,
including the liquid crystalline contribution to the fluid stress
tensor.

The essential physics is that flow tends to align the rodlike
molecules, typically roughly parallel to the flow direction, and hence
stabilizes a nematic, or aligned, state. To study other complex fluids
we would have a structural variable analogous to $\boldsymbol{Q}$;
\emph{e.g.} in the wormlike micelle system we might need, in addition
to the orientation tensor, the dynamics of the mean micellar size.  
We have augmented the Doi model by allowing for concentration diffusion
driven by chemical potential gradients, included the dynamical
response to inhomogeneities in liquid crystalline order and
concentration, and included the translational entropy of mixing which
gives the system a biphasic coexistence regime in the absence of flow.
For a given stress or strain rate we determine phase coexistence by
explicitly constructing a stable coexisting steady state, which require
inhomogeneous terms in the equations of motion (arising here from free
energy terms which penalize inhomogeneities in $\boldsymbol{Q}$ and
$\phi$). This procedure and the model have been documented elsewhere
\cite{olmsted92,olmstedlu97}, and the interface construction will be
discussed in more detail in Sec.~\ref{sec:interface}.

To calculate the phase diagram we solve for the steady state
homogeneous solutions to the coupled dynamical equations for
$\{\phi,\boldsymbol{Q},{\bf V}\}$.  This yields a set of solutions
which are then candidates for phase coexistence. Coexistence is
possible with either common stress or common strain rate in the
coexisting phases, depending on geometry, and must be examined for all
pairs of stable homogeneous states.  We expect coexisting states to
have different values for $\phi$, $\boldsymbol{Q}$, and either the
stress or strain rate. As mentioned above, we determine coexistence by
finding the locus of control parameters for which a stable interfacial
solution between two homogeneous states exists. We parametrize the
shear stress and strain rate as
\begin{eqnarray}
\widehat{\dot{\gamma}} &=& \frac{\dot{\gamma} L^2}{6D_{\mit r0}\nu_1\nu_2^2} \\
\widehat{{\sigma}_{xy}} &=&\frac{\sigma_{xy}\nu_2 L^3}{3
  k_{\scriptscriptstyle B} T},
\end{eqnarray}
where $D_{\mit r0}$ is the rotational diffusion coefficient, and
$\nu_1$ and $\nu_2$ are ${\cal O\/}(1)$ geometric constants.

The Doi model has three stable steady states in shear flow
\cite{bhave93}: A weakly-ordered \emph{paranematic} state {\sf I},
with the major axis of the order parameter in the shear plane; a
\emph{flow-aligning} state \textsf{N}, with a larger order parameter
and major axis in the shear plane; and a \emph{log-rolling} state
\textsf{L}, with major axis in the vorticity direction.
Fig.~\ref{fig:constit} shows homogeneous constitutive relations for
the \textsf{I} and \textsf{N} states. As can be seen, the \textsf{N}
and \textsf{L} states are, successively less viscous than the
\textsf{I} phase at the same concentration, with a viscosity which
decreases slightly with increasing concentration (reflecting the
greater order and hence lower viscosity of more concentrated phases),
in contrast to the less-ordered \textsf{I} phases, whose viscosity
increases with concentration, as is usual for colloidal suspensions.
{\begin{figure}
\hsize\columnwidth\global\linewidth\columnwidth
\displaywidth\columnwidth
\epsfxsize=3.5truein
\centerline{\epsfbox[70 230 670 650]{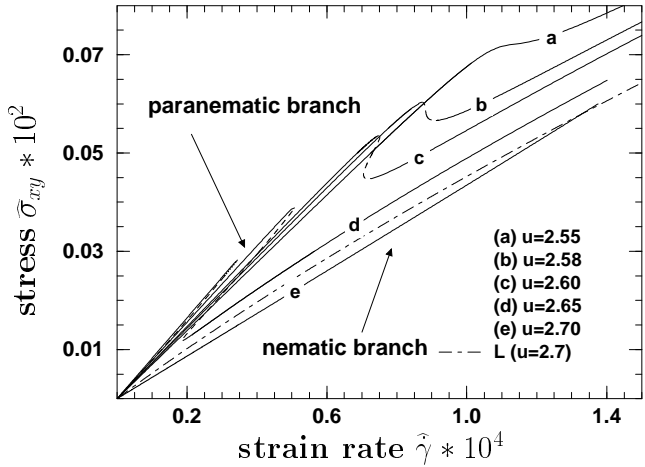}}
\caption{Homogeneous stress $\hat{\sigma}_{xy}$ vs. strain rate
  $\widehat{\dot{\gamma}}$ curves for various excluded volumes $u$
  ($u$ is proportional to $\phi$) and $L=5$. Shown are the \textsf{I}
  and \textsf{N} branches. The \textsf{L} branch is only stabilized
  for the high concentration ($u=2.7$).}
\label{fig:constit}
\end{figure}}
\subsection{Common Stress Phase Separation}
We first discuss the phase diagrams for common stress coexistence, in
which the phase separation is radial in a cylindrical Couette flow.  For common
stress coexistence of two phases $I$ and $II$, the fraction $\zeta$ in
phase $I$ is determined by the lever rule,
\begin{subequations}
  \label{eq:lever}
  \begin{eqnarray}
    \bar{\phi} &=& \zeta \phi_{\scriptscriptstyle I} + (1-\zeta)
    \phi_{\scriptscriptstyle II} \label{eq:lev1}\\
    \bar{\dot{\gamma}} &=& \zeta \dot{\gamma}_{\scriptscriptstyle I} 
    + (1-\zeta) \dot{\gamma}_{\scriptscriptstyle II},\label{eq:lev2}
  \end{eqnarray}
\end{subequations}
where $\bar{\phi}$ and $\bar{\dot{\gamma}}$ are mean values.
Fig.~\ref{fig:tie} shows the phase diagram calculated for \textsf{I-N}
coexistence for $L=5$. The tie lines denoting pairs of coexisting
phases are horizontal in the $\sigma_{xy}\!-\!u$ plane, and have
positive slopes in the $(\widehat{\dot{\gamma}}\!-\!u)$ plane because
the more concentrated nematic phase flows faster at a given stress.
For weak stresses the equilibrium system is slightly perturbed and the
tie lines are almost horizontal, while at high stresses the tie lines
become steeper as the composition difference between the phases
decreases and vanishes at a critical point.

{\begin{figure}
\hsize\columnwidth\global\linewidth\columnwidth
\displaywidth\columnwidth
\epsfxsize=5.0truein
\centerline{\epsfbox[130 230 800 650]{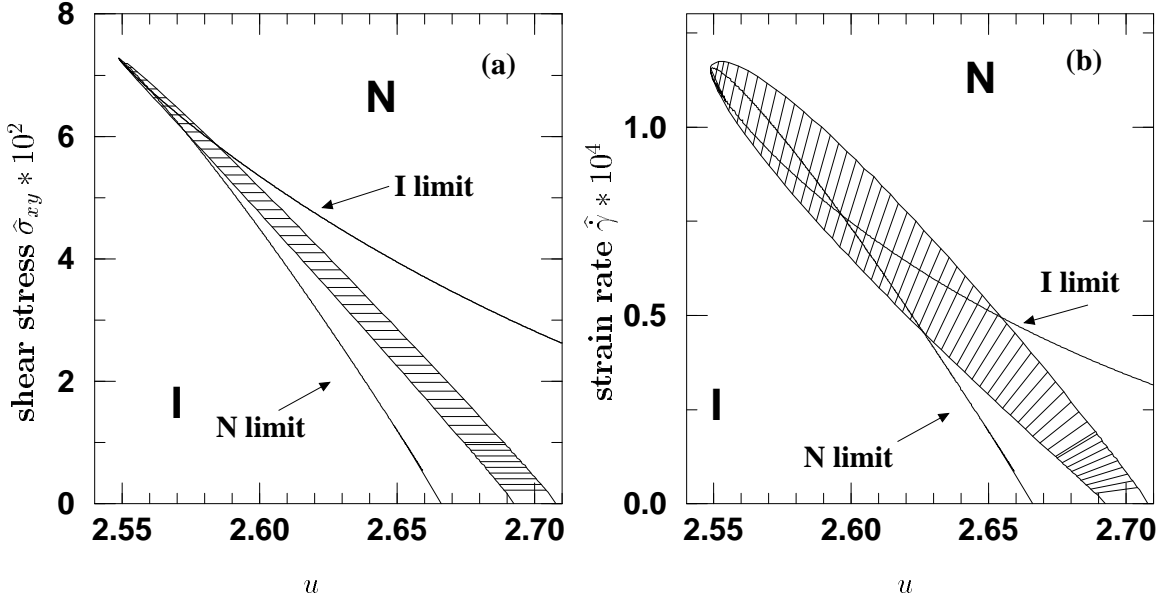}}
\caption{
  Phase diagram in the $(\hat{\sigma}_{xy}\!-\!u)$ (a) and
  $(\widehat{\dot{\gamma}}\!-\!u)$ (b) planes, along with limits of
  stability of \textsf{I} and \textsf{N} phases. Tie lines (horizontal
  in (a), sloped in (b) connect coexisting phases.}
\label{fig:tie}
\end{figure}}

{\begin{figure}
\displaywidth\columnwidth
\epsfxsize=\displaywidth
\centerline{\epsfbox[100 220 1400 630 ]{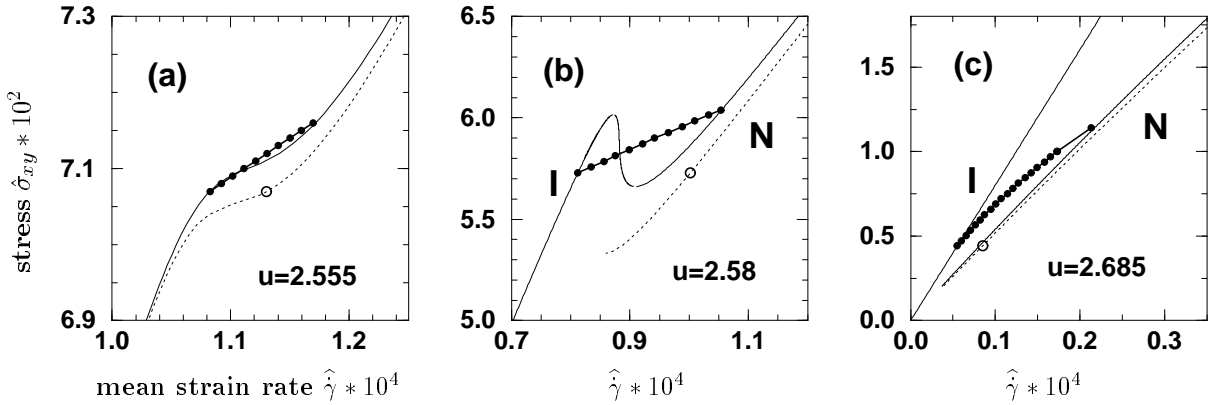}}
\caption{Stress $\sigma_{xy}$ vs. mean strain rate $\bar{\dot{\gamma}}$ 
  for common stress coexistence.  Solid lines are constitutive curves
  for the \textsf{I} and \textsf{N} branches; dotted lines and
  $\boldsymbol{\circ}$ denote the \textsf{N} branch with which each
  \textsf{I} branch coexists at the low strain rate coexistence
  boundary. $\bullet$ denotes banded stresses, whose plateaus do
  \emph{not} satisfy an equal area construction with the homogeneous
  constitutive curve.}
\label{fig:stress-strainbar0}
\end{figure}}

\noindent{\textbf{Mean Constitutive Relations---}}From the information
in Fig.~\ref{fig:tie}b we can calculated the mean constitutive
relation that would be measured in an experiment on a system with a
given prescribed mean concentration. Upon applying stress at a given
concentration, the system traces a vertical path through
Fig.~\ref{fig:tie} until the two-phase region is reached, during which
$\hat{\sigma}_{xy}(\bar{\dot{\gamma}})$ varies smoothly. At this
stress a tiny band of \textsf{N} phase develops, with composition and
strain rate determined by the lever rule; and
$\hat{\sigma}_{xy}(\bar{\dot{\gamma}})$ is non-analytic
(Fig.~\ref{fig:stress-strainbar0}), exhibiting a change in slope. As
the stress and, hence the mean strain rate, increases further the
system visits successive tie lines in the $\dot{\gamma}\!-\!u$
plane, each with a higher stress and mean strain rate and different
coexisting concentrations.  For $\bar{\phi}$ close to the equilibrium
\textsf{I-N} transition (Fig.~\ref{fig:stress-strainbar0}c) the tie
lines in the $(\widehat{\dot{\gamma}}\!-\!u)$ plane are fairly flat
and the stress $\sigma_{xy}$ changes significantly through the
two-phase region.  More dilute systems
(Fig.~\ref{fig:stress-strainbar0}a,b) have steeper tie lines and
straighter and flatter `plateaus' in
$\hat{\sigma}_{xy}(\bar{\dot{\gamma}})$.

Controlled strain rate experiments should follow the homogenous flow
curves, except for the coexistence regime. In this case, analogy with
equilibrium systems suggests that the system should eventually
nucleate into a phase-separated banded state, with a corresponding
stress change. Experiments on wormlike micelles display this kind of
behavior upon increasing the strain rate above that of the phase
boundary. If the mean strain rate is on an unstable part of the flow
curve, we expect a `spinodal' (or mechanical) instability.  There is a
small region (inside the loop in Fig.~\ref{fig:tie}a) where the system
is unstable when brought, at controlled strain rate, into this region
from either the \textsf{I} or \textsf{N} states. This corresponds to
constitutive curves with the shape of curve \textsf{b} in
Fig.~\ref{fig:constit}.

For controlled stress experiments, for stresses larger than the
minimum coexistence stress and less than the local maximum we expect
the system to follow the homogenous flow curve until a nucleation
event occurs (Fig.~\ref{fig:stress-strainbar0}b).  Then, the strain
rate should increase to either that of proper banded or single
\textsf{N} phase strain rate, depending on the magnitude of the
stress. For stresses larger than the \textsf{I} limit of stability we
expect, again by analogy with equilibrium, a spinodal-type
instability. This simple picture is not quite corroborated in wormlike
micelles: Grand {\em et al.\/} reported that a stress within a narrow
range above the coexistence stress could be applied, and the system
remained on the ``metastable'' branch indefinitely. 

Analogy with equilibrium systems suggests similar behavior upon
reducing the stress or strain rate from the high-shear branch. 
Careful experiments (on any system) are needed to test the idea. 
For example, it is interesting to examine that,
upon reducing the strain rate below the upper strain rate for the
onset of shear banding and above the limit of stability of the high
strain rate branch, whether the stress would spontaneously increase into the
banded state.

{\begin{figure}
\hsize\columnwidth\global\linewidth\columnwidth
\displaywidth\columnwidth
\epsfxsize=5.5truein
\centerline{\epsfbox[70 220 740 650]{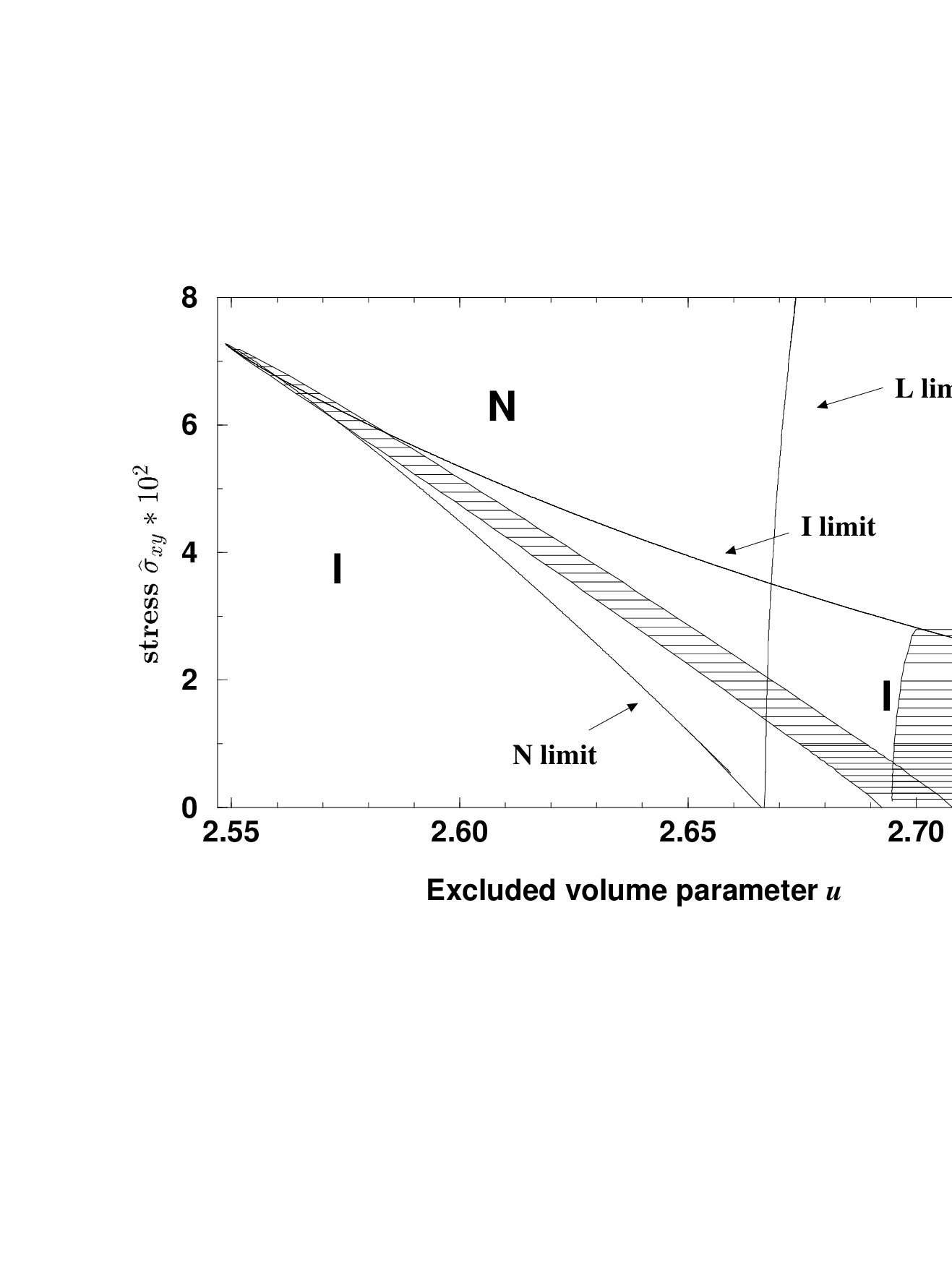}}
\caption{
  Composite phase diagrams for {\sf I-L} and {\sf I-N} coexistence at
  common stress.  This represents \emph{two} overlayed phase diagrams,
  and not a single phase diagram.}
\label{fig:tieboth}
\end{figure}}
\noindent{\textbf{Log Rolling Phase---}}
Fig.~\ref{fig:tieboth} shows the phase diagram for paranematic-log
rolling \textsf{I-L} coexistence. For non-zero stress the biphasic
region shifts to higher concentrations, since the stability limit of
the {\sf L} phase shifts to higher concentrations.  Since the {\sf I}
and {\sf L} phases have major axes of alignment in orthogonal
directions, there is not a critical point; rather, the biphasic region
ends when the {\sf I} phase becomes unstable to the {\sf N} phase.  We
have also computed \textsf{N-L} phase coexistence, but cannot resolve
this (very concentrated, $u\simeq 3$) regime accurately and do not
present these results here.

Can one observe {\sf I-L} coexistence? This can only occur for
concentrations above that necessary for equilibrium phase separation.
One could conceivably prepare an equilibrium \textsf{I-N} mixture with
the nematic phase in the log-rolling geometry.  Upon applying shear,
the system would then maintain coexistence and move through the {\sf
  I-L} two-phase region.  However, the \textsf{I} phase is, itself,
within the two phase region for \textsf{I-N} phase separation, so we
expect the prepared coexisting \textsf{I-L} state to be metastable
under shear.

\subsection{Common Strain Rate Phase Separation}
{\begin{figure}
\hsize\columnwidth\global\linewidth\columnwidth
\displaywidth\columnwidth 
\epsfxsize=6.2truein
\centerline{{\epsfbox[100 220 880 650]{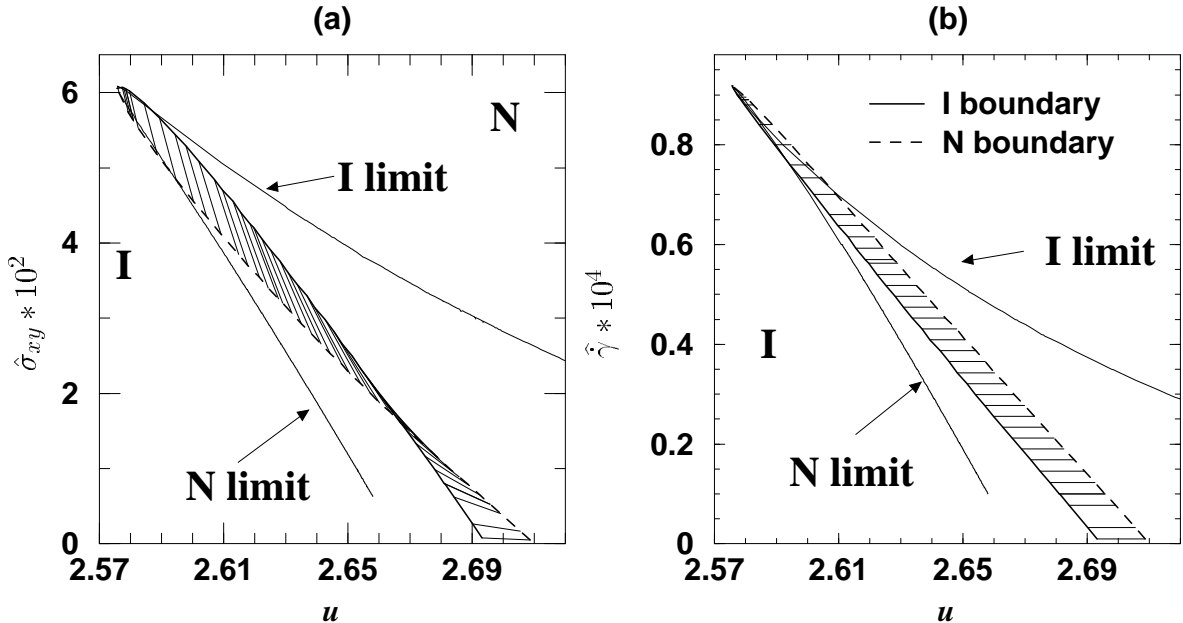}}}
\caption{
  Common strain rate phase diagram in the
  $(\widehat{\dot{\gamma}}\!-\!u)$ (a) and $(\hat{\sigma}_{xy}\!-\!u)$
  (b) planes. The solid and dashed lines denote the phase boundaries
  for, respectively, the \textsf{I} and \textsf{N} phases.}
\label{fig:ties}
\end{figure}}

Common strain rate phase separation can be calculated exactly
analogous to that for common stress phase separation. The resulting
phase diagram for \textsf{I-N} coexistence, for $L=5$, is shown in
Fig.~\ref{fig:ties}. The shear stress and composition are partitioned
according to the lever rule in Fig.~\ref{fig:ties}b, with
\begin{equation}
  \label{eq:8}
      \bar \sigma_{xy} = \zeta {\sigma}_{xy\scriptscriptstyle I} 
    + (1-\zeta) {\sigma}_{xy\scriptscriptstyle II},
\end{equation}
In this case tie lines connecting coexisting phases are parallel in
the $\dot{\gamma}\!-\!u$ plane, and have a negative slope in the
$(\hat{\sigma}_{xy}\!-\!u)$ plane because the {\sf I} phase coexists
with a denser and less viscous \textsf{N} phase.  There is an
interesting crossover in the $(\hat{\sigma}_{xy}\!-\!u)$ plane.  For
dilute systems the stress in the \textsf{N} phase immediately outside
the biphasic regime is {\sl less\/} than the stress just before the
system enters the biphasic region (Fig.~\ref{fig:ties}b).  Since the
stress of the {\sf N} branch is less than that of the {\sf I} branch
at the same strain rate and composition, we expect a decrease in the
stress across the biphasic regime if composition effects are weak,
\emph{e.g.} near a critical point.  For higher mean compositions the
stress \emph{increases} across the biphasic regime, because the width
of the biphasic regime overcomes the shear thinning effect.

\noindent{\textbf{Mean Constitutive Relations---}} The mean
constitutive relation that could be measured in an experiment may be
calculated from Fig.~\ref{fig:ties}a, and are shown in
Fig.~\ref{fig:stressstrainbars0}. At higher concentrations the plateau
has a positive slope while, coinciding with the crossover noted above,
for lower concentrations the plateau has a negative slope.  A negative
slope usually signifies a bulk instability, but here each band lies on
a stable branch of its particular constitutive curve and the flow
should be stable. Stable `negative-slope' behavior has been seen in
shear-thickening systems which phase separate at common stress
\cite{boltenhagen97a,boltenhagen97b}, although in that case the 
mean constitutive curve was different, consisting of a backwards
\textsl{S} curve and non-monotonic behavior only with multiple
stresses for a given strain rate.

{\begin{figure}
\displaywidth\columnwidth
\epsfxsize=6.5truein
\centerline{\epsfbox[80 210 1300 600]{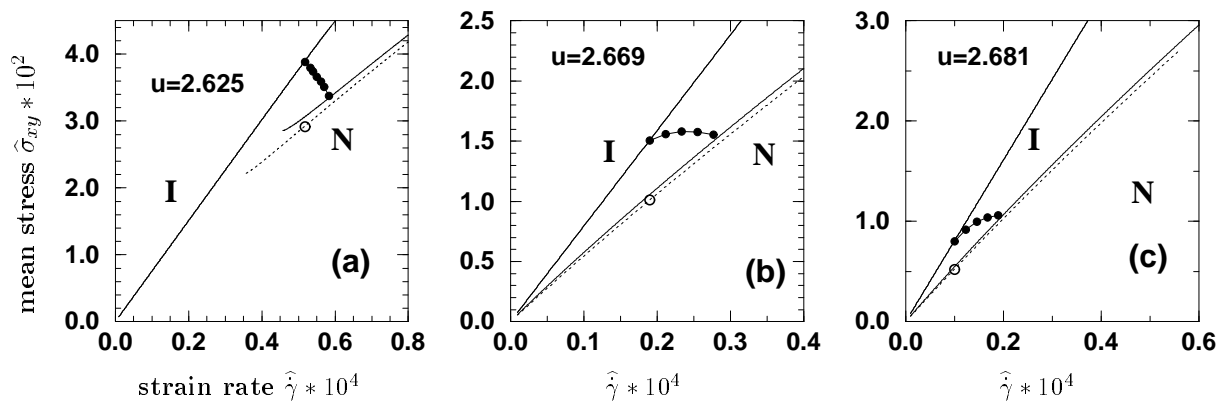}}
\caption{Stress--strain-rate $\bar{\sigma}_{xy}(\hat{\dot{\gamma}})$
  curves for common strain rate coexistence.  Same notation as
  Fig.~\ref{fig:stress-strainbar0}.}
\label{fig:stressstrainbars0}
\end{figure}}

Based on analogies with equilibrium, we naively expect controlled
strain rate and controlled stress experiments on concentrations such
as those in Fig.~\ref{fig:stressstrainbars0}a-b to yield behavior
similar to that for common stress phase separation, with nucleated or
spinodal behavior depending on the applied strain rate, and the same
caveat applying to decreasing the strain rate from agove. The
situation for compositions with curves such as
Fig.~\ref{fig:stressstrainbars0}a is qualitatively different. Here
there is a range of stresses with \emph{three} stable states:
homogeneous \textsf{I} and \textsf{N} branches, and a banded
intermediate branch.  For controlled stress experiments, one
possibility is that the \textsf{I} and \textsf{N} branches are favored
in their respective domains of stresses. For example, in start-up
experiments the system would remain on the \textsf{I} branch until a
certain stress, at which point it would nucleate after some time. If
the system nucleated onto the coexistence branch, increasing the
stress further would return the system to the \textsf{I} branch.
Since it nucleated \emph{from} the \textsf{I} branch, it is more
likely to jump directly to the \textsf{N} branch.  Similar behavior is
to be expected upon reducing the stress from the \textsf{N} phase.
Another possibility is intrinsic hysteresis: that is, the system never
jumps until reaching its limit of stability (from either the
\textsf{I} or \textsf{N} side).  The present theory cannot address
this question. For controlled strain rate experiments, it could, in
principle, be possible to maintain a stress on the two state region,
although in practice this would also seem to be quite difficult, and
would seem to be mechanically unstable.  In the case where stable
composite curves with negative slopes were accessed, stress was the
control variable \cite{HBP98}.
  
\subsection{Common Stress or Common Strain Rate?}
What about the relative stability of phase separation at common stress
or strain rate?  While our one-dimensonal calculations cannot address
this question, we have examined the two phase diagrams in the
$\sigma_{xy}\!-\!\mu$ and $\dot{\gamma}\!-\!\mu$ planes, where $\mu$
is the chemical potential \cite{olmstedlu97}.  This can be seen in
Fig.~\ref{fig:resolve}a-b where, for example, the \textsf{I} boundary
for common strain rate phase separation ($I_{\gamma}$) lies in the
\textsf{N} region for common stress phse separation, in the
$\mu\!-\!\sigma_{xy}$ plane.  This occurs because the stress of the
\textsf{I} phase, at common strain rate, is larger than the stress of
the \textsf{N} phase, due to the shear thinning nature of the
transition.  Conversely, the \textsf{I} phase at common stress lies
within the \textsf{I} region of the common strain rate phase diagram.
Analogy with equilibrium phase transitions suggests that, since the
\textsf{I} phase of common strain rate phase separation thus lies on
the ``wrong'' side of the phase boundary for common stress, given by
the line in the $\mu\!-\!\sigma_{xy}$ plane, it would be unstable (or
metastable) to phase separation at common stress. Conversely, the
\textsf{I} phase at common stress is on the ``correct'' side of the
coexistence line in the $\mu\!-\!\dot{\gamma}$ plane, and, again based
on analogy with equilibrium, might be expected to be stable. Note that
if the transition were shear thickening the situation would be
reversed, and the arguments above would lead to common stress phase
separation being unstable (or metastable) with respect to common
strain rate phase separation.

Boundary conditions may also play a role.  In a Couette device the
slight inhomogeneity of Couette flow induces an asymmetry between the
inner and outer cylinders, exactly the symmetry of common stress phase
separation (Fig.~\ref{fig:both}).  This should enhance the stability
of common stress phase separation.  Cone-and-plate rheometry induces a
similar preference for the common stress geometry.

{\begin{figure}
\displaywidth\columnwidth
\epsfxsize=6.5truein
\centerline{\epsfbox[26 102 571 410]{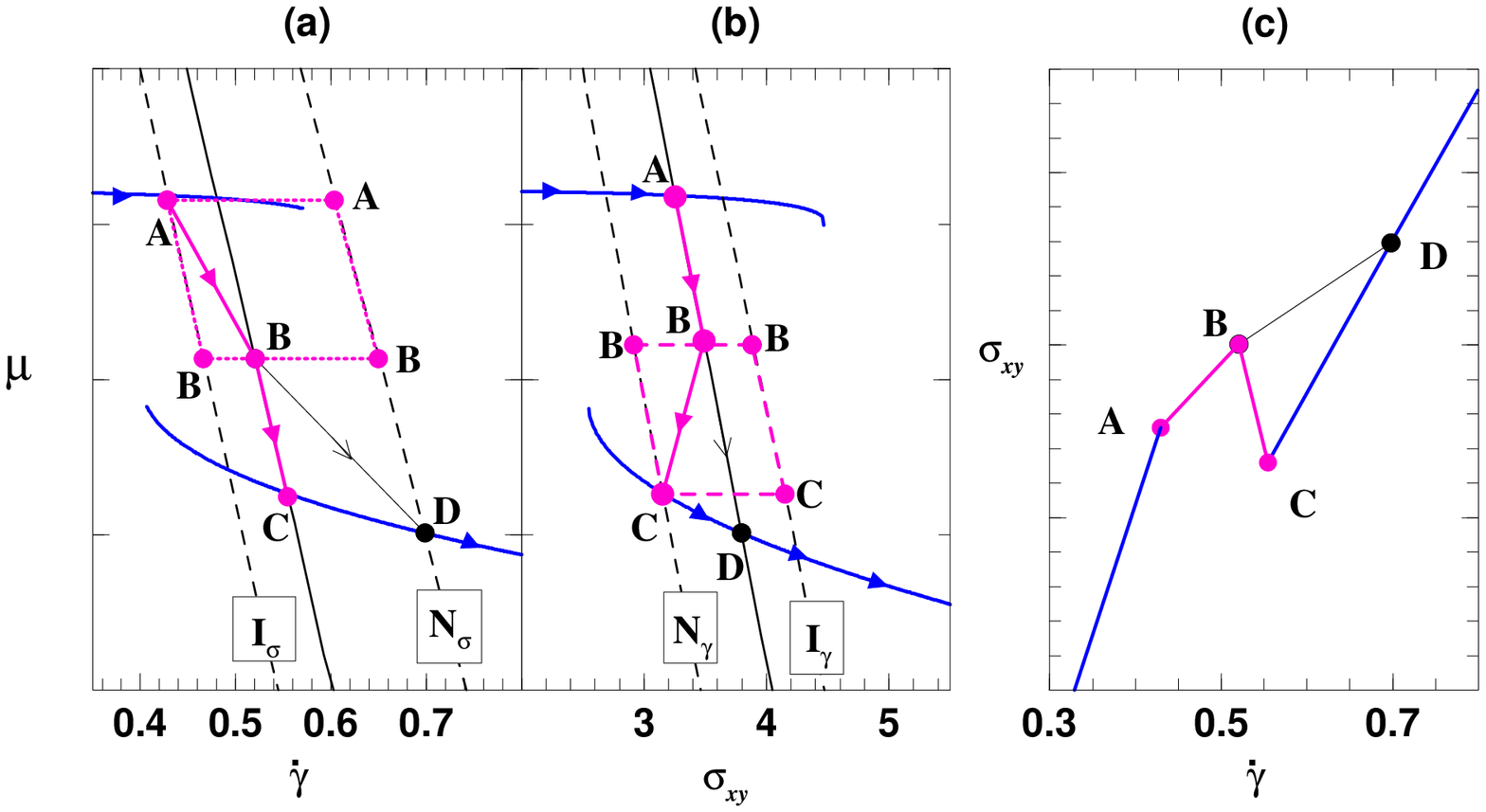}}
\caption{Phase diagrams in the (a) $\mu\!-\!\dot{\gamma}$ and (b)
  $\mu\!-\!\sigma_{xy}$ planes for \textsl{I-N} coexistence (the
  \textsl{N} state is stable for higher strain rate or stress,
  respectively). The thin vertical solid lines denote phase
  coexistence at common strain rate and stress in (a) and (b),
  respectively.  The broken lines marked $I_\gamma$ and $N_\gamma$
  denote the coexisting states at common strain rate, in the
  $\mu\!-\!\sigma_{xy}$ plane (b); while the broken lines $I_\sigma$
  and $N_\sigma$, denote the coexisting states at common stress, in
  the $\mu\!-\!\dot{\gamma}$ plane. (c) is the mean stress vs. strain
  rate curve. ABD denotes a path in the phase diagrams for common
  stress phase separation, while ABCD is a path which switches from
  common stress phase separation to common strain rate phase
  separation at B.  }
\label{fig:resolve}
\end{figure}}

An alternative possibility is presented in Fig.~\ref{fig:resolve}.  If
one argues that, in steady state, among the possible phases which are
compatible with the interface solvability condition, the chemical
potential reaches its minima so that no more diffusive material flux
is possible.  Based on such a criterion, upon increasing the strain
rate for a given mean concentration the stable phase is that with the
lowest chemical potential. The thick horizontal arrows in
Fig.~\ref{fig:resolve} denote the $\mu(\sigma_{xy})$ and
$\mu(\dot{\gamma})$ paths for the homogeneous high and low shear rate
states, in the two phase diagrams.  The \textsf{I} branch becomes
unstable at A to phase separation at common stress, when the
homogeneous path first crosses the phase boundary in the
$\mu\!-\!\sigma$ plane. For higher stresses the system follows the
segment AB in Fig.~\ref{fig:resolve}b, along the phase coexistence
line at common stress, and follows the stress plateau AB in
Fig.~\ref{fig:resolve}c.  In the $\mu-\dot{\gamma}$ plane the system
phase separates, and the chemical potential as a function of mean
strain rate follows the diagonal path AB in Fig.~\ref{fig:resolve}a
(the dotted lines denote the strain rates of the coexisting phases).

Upon increasing the strain rate further than point B, the chemical potential 
of the system can decrease by phase separating at a
common strain rate. This reduces the chemical potential, at a given
strain rate, from that of the segment BD to that of segment BC. Hence
the system would take the path BC along the phase boundary in the
$\mu\!-\!\dot{\gamma}$ plane, as far as point C, upon which the phase
boundary crosses the homogeneous curve for the high strain rate phase
of the given mean concentration. The path would be the diagonal path
BC in the $\mu\!-\!\sigma_{xy}$ plane, and would correspond to the
negative-sloped segment BC in the flow curve, Fig.~\ref{fig:resolve}c.
Finally the system follows the high strain rate branch, through CD.

Upon increasing the controlled stress, the system would be expected to
follow ABD. Upon decreasing the stress or the strain rate, DC-bottom
jumping is expected.  These scenarios follow from minimizing the
chemical potential subject to the solvability constraint should phases
coexist.  Its correctness, of course, should be further examined by
the full time evolution of the original dynamic equations.

\noindent{\textbf{Experimental Studies---}}
Mather \textit{et al.} \cite{mather97} studied a liquid crystalline
polymer melt (an aromatic polyester) and determined the lower limit of
the \textsf{I-N} phase boundary in the $\dot{\gamma}\!-\!T$ plane.
The studies most relevant to the Doi model for rigid rod suspensions
have been on wormlike micellar solutions near their isotropic-nematic
coexistence region \cite{berret94a,schmitt94}, where common-stress
banding was observed with a plateau stress that became steeper for
concentrations closer to the equilibrium \textsf{I-N} phase boundary,
in qualitative agreement with our results. Common strain rate banding
has not bee seen in these systems.  Micelles are considerably more
complicated than simple rigid rods, because they are not strictly
rigid and their length (and hence coupling to flow) is a strong
dynamic function of concentration.  Experiments on micelles far from
an apparent nematic transition exhibit common stress shear banding
with nearly flat coexistence plateaus
\cite{berret94a,grand97,Berr97,rehage91,BritCall97}, consistent with a
concentration-independent \cite{cates90} instability (or transition).
In kinetics studies the delay time before the transition to a banded
(or high strain rate) flow in controlled stress start-up `quenches'
diverged for a window of stresses slightly above the banding stress,
whereas controlled strain rate `quenches' always decayed, eventually,
onto a banded flow state.  These interesting behaviors cannot be
explained by the topologies of the phase diagrams in
Fig.~\ref{fig:tie}.  Bonn \emph{et al.}  \cite{Bonn+98} recently
studied lamellar surfactant systems and observed slowly coarsening
bands in the common strain rate geometry; and for controlled strain
rate measurements they found transient constitutive curves analogous
to Fig.~\ref{fig:stressstrainbars0}a or~\ref{fig:stressstrainbars0}b,
consistent with common-strain rate phase separation.  The true steady
state behavior was not measured.
\section{Interface Construction}\label{sec:interface}
Several microscopic and phenomenological models, as well as the
apparent underlying flow curves for wormlike micelles, show an
apparent degeneracy in the shear stress at which coexistence occurs
(in the case of coexistence at a common stress). To resolve this
degeneracy we have relied on the presence of inhomogeneous terms in
the dynamical equations of motion, and determined the selected stress
as that stress which allows a stable interfacial solution.  In this
section we explore this in more detail using a toy constitutive model.
Similar arguments were given Spenley \emph{et al.}  \cite{spenley96}
in a different languauge, and in a recent more rigorous study
\cite{jsplanar}.

We consider planar flow with a velocity field ${\bf v}({\bf
  r})=v(y){\bf\hat{x}}$, with $\dot{\gamma}(y)\equiv\partial
v/\partial y$, and postulate the following constitutive relation for
the shear stress:
\begin{equation}
  \label{eq:1}
  \sigma(\dot{\gamma}) = \sigma_h(\dot{\gamma}) - D(\dot{\gamma})\,
  \partial_y^2\dot{\gamma}.
\end{equation}
The homogeneous flow curve $\sigma_h(\dot{\gamma})$ is non-monotonic,
as in Fig.~\ref{fig:js}; and can be derived for a system with an
underlying transition, as for the modified Doi model above, or from
phenomenological models, such as the widely-used Johnson-Segalman (JS)
model \cite{johnson77}.  Gradient terms may come from the diffusion of the 
stress elements \cite{elkareh89}. The flow curve shown in
Fig.~\ref{fig:js} is for the JS model. In the (some what artificial)
model where only the shear stress diffusion is considered, the steady
flow condition for the JS model has the form of Eq.~(\ref{eq:1}), with
$D(\dot{\gamma})\propto 1/(1+\dot{\gamma}^2)$.
{\begin{figure}
\hsize\columnwidth\global\linewidth\columnwidth \epsfxsize=3.5truein
\centerline{\epsfbox[100 230 700 660]{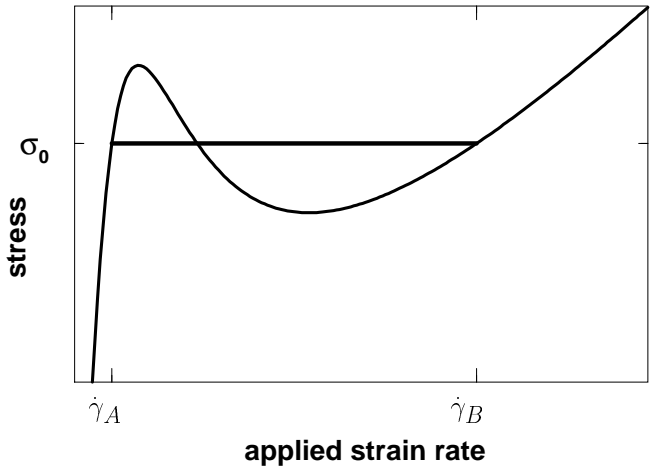}}
\caption{Non-monotonic homogeneous flow curve
  $\sigma_h(\dot{\gamma})$ for the Johnson-Segalmann model. The thick
  curve denotes a banded flow between strain rates
  $\dot{\gamma}_{\scriptscriptstyle A}$ and
  $\dot{\gamma}_{\scriptscriptstyle B}$ at stress $\sigma_0$.}
\label{fig:js}
\end{figure}}

The steady state condition for planar flow is a uniform shear stress,
\begin{equation}
  \label{eq:2}
  \sigma_0 = \sigma_h(\dot{\gamma}) - D(\dot{\gamma})\,
  \partial_y^2\dot{\gamma},
\end{equation}
with $\sigma_0$ a constant. In an infinite system, an interfacial shear
banding solution at a given stress $\sigma_0$ satisfies
Eq.~(\ref{eq:2}), with boundary conditions
\begin{subequations}
  \label{eq:3}
  \begin{eqnarray}
    \dot{\gamma}(-\infty) &=& \dot{\gamma}_A \\
    \dot{\gamma}(\infty) &=& \dot{\gamma}_B.
  \end{eqnarray}
\end{subequations}
\begin{equation}
\partial_y \dot{\gamma}(\pm \infty) =0
\end{equation}
Hence, given the
second-order differential equation, the system is overdetermined. A
solution is only possibly when these two conditions coincide, which
may be obtained by varying the stress $\sigma_0$.  It is
straightforward to integrate Eq.~(\ref{eq:2}) to show that a solution
is possible when $\sigma_0$ satisfies the following condition
\cite{jsplanar}
\begin{equation}
  \label{eq:7}
  \int_{\dot{\gamma}_{\scriptscriptstyle 
      A}}^{\dot{\gamma}_{\scriptscriptstyle A}} \frac{\sigma_0-
      \sigma_h(\dot{\gamma})}{D(\dot{\gamma})} = 0.
\end{equation}
Note that this is not an equal areas construction, unless
$D(\dot{\gamma})$ is a constant $D$.

Further insight may be obtained by casting the interface solution in
terms of a dynamical system.  Defining
\begin{subequations}
  \begin{eqnarray}
    \label{eq:4}
    p&=& \dot{\gamma} \\
    q&=& \partial_y p\equiv p',
  \end{eqnarray}
\end{subequations}
Eq.~(\ref{eq:2}) becomes the following dynamical system, with $y$
playing the role of time.
\begin{subequations}
  \label{eq:5}
\begin{eqnarray}
  p'&=&q\\
  q'&=& \frac{\sigma_0 - \sigma_h(p)}{D(p)}.
\end{eqnarray}
\end{subequations}
For $\sigma_0$ within the non-monatonic region of the flow curve the
system has three fixed points $p_{\ast}=\left\{p_{\scriptscriptstyle
    A},p_{\scriptscriptstyle B}, p_{\scriptscriptstyle C}\right\}$ on
the axis $q=0$, corresponding to the strain rates of the three homogeneous flows.
Linear stability analysis yields the
stable and unstable manifolds of points $A$ and $B$, with eigenvalues
\begin{equation}
  \label{eq:6}
  \lambda_{\pm} = \pm\sqrt{\left.\left[\frac{1}{D(p)} 
      \frac{d\sigma_h}{dp}\right]\right|_{p=p_{\ast}}}
\end{equation}
and eigenvectors at angles $\theta=\arctan \lambda$ with respect to
the $p$-axis. Point $C$ has imaginary eigenvalues and is a cycle,
while $A$ and $B$ are saddles with stable and unstable directions.

An interfacial solution corresponds to an orbit connecting
saddles $A$ and $B$, and is denoted a saddle connection; it it also
called a heteroclinic orbit, since it connects two different fixed
points.  This set of ordinary differential equations (ODE) does not
generally have a saddle connection for an arbitrary $\sigma_0$ in the
multi-valued region. It can be shown \cite{jsplanar} that for models (with arbitrary
numbers of dynamical variables) in planar shear flow with differential
non-local terms that, apart from accidents, a saddle connection only
exists at isolated points in the control parameter
space. Here, the control parameters are $\sigma_0$ and parameters
which change the shape of $\sigma_h$ \cite{jsplanar,spenley96}, while
for the Doi model above, the control parameters are (for a given set of
molecular parameters such as $L$) $\mu$ and $\sigma_{xy}$ for common
stress phase separation; and $\mu$ and $\dot{\gamma}$ for common
strain rate phase separation.
{\begin{figure}
\hsize\columnwidth\global\linewidth\columnwidth
\epsfxsize=6.2truein
\centerline{{\epsfbox[72 290 540 500]{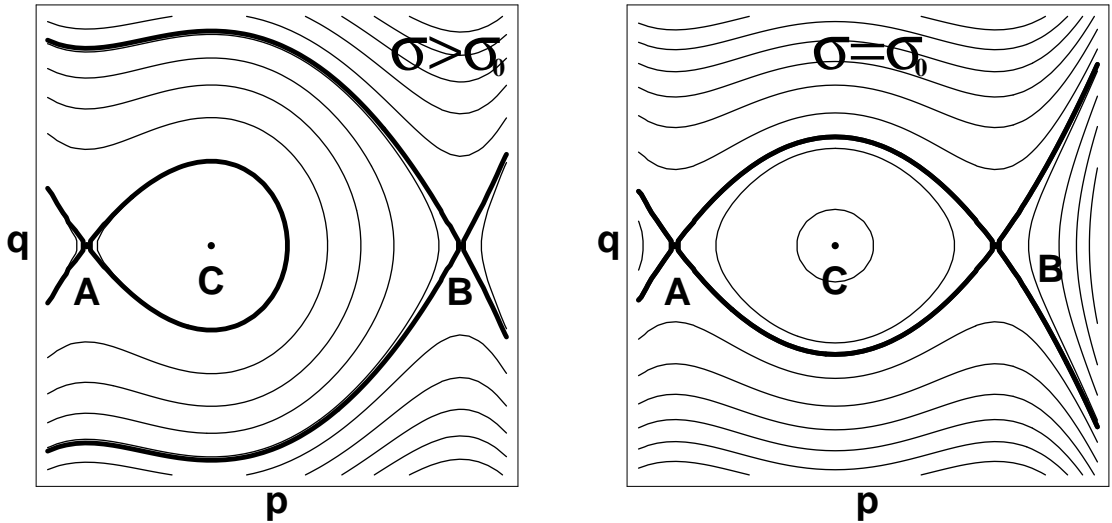}}}
\caption{Orbits in the phase space for the dynamical system of
  Eqs.~(\ref{eq:5}). (Left) Away from the coexistence stress,
  $\sigma>\sigma_0$, a saddle connection does not exist and $A$ is
  connected to itself by a homoclinic orbit. [A similar situation
  exists for $\sigma<\sigma_0$, with the homoclinic orbit returning to
  $B$ instead of $A$.] (Right) At the coexistence stress
  $\sigma=\sigma_0$ a heteroclinic orbit connecting $A$ and $B$
  (saddle connection) exists. The fixed points are at $q=0$, and the
  solid lines show orbits beginning and ending on the stable fixed
  points $A$ and $B$.}
\label{fig:phase}
\end{figure}}

Fig.~\ref{fig:phase} shows the evolution of ``orbits'' in the $p-q$
phase space as the stress is tuned. For $\sigma=\sigma_0$ a  
heteroclinic orbit exists, connecting $A$ and $B$. This corresponds to
an elementary shear band solution, in which one portion of the sample
lies on the high strain rate branch $B$, another portion lies on the
low strain rate branch $A$, and a single interface separates the two
phases. For $\sigma\neq\sigma_0$ there is no heteroclinic orbit or
saddle connection, and hence no stationary interface.
Fig.~\ref{fig:phase} (Left) shows a stress slightly greater than $\sigma_0$, 
where a homoclinic orbit connecting state $A$ to itself.
Kramer \cite{kramer81a} pointed out in the context of reaction
diffusion equations that such a homoclinic orbit corresponds to the
critical droplet in a metastable phase of $A$ material. Note that, although in
real space it goes from $A$ at $y=-\infty$ to $A$ at $y=+\infty$, the dominant
spatial variation is in fact localized, with a size that vanishes when
the stress reaches the maximum of the flow curve in Fig.~\ref{fig:js}
(at which the fixed points $A$ and $C$ annihilate).  Slightly larger
droplets are unstable and, when the full dynamics are returned to the
problem, presumably flow to the high strain rate branch $B$, while
smaller droplets are expected to decay back to $A$.  By analogy with
equilibrium behavior, for $\sigma>\sigma_0$ we expect phase $B$ to be
the long time steady state, if fluctuations (\emph{i.e.} noise,
thermal or otherwise) were included.

\section{Conclusion}
We have outlined the phenomenology of phase separation of rigid rod
suspensions in shear flow, using the modified Doi model.  Phase
separation may  occur with  common stress
\emph{or} strain rate, corresponding to the different coexistence geometry.  We have
calculated coexistence among three phases (paranematic, flow-aligning
nematic, and log-rolling), while only two equilibrium phases exist.
That is, the full rotational symmetry of an equilibrium nematic is
broken by the biaxial shear flow, leaving two possible stable nematic
orientations (the in-plane \textsf{I} and \textsf{N} states, and the
out of plane \textsf{L} state).  The shear thinning nature of the
transition suggests that common stress phase separation is stable;
while appealing to a minimization of the chemical potential, subjected to the interface solvability,  predicts a curious
crossover from common stress to common strain rate phase separation.
We do not know which of these, or other, possibilities, are the
physical ones. The composite stress strain curves depend on the
coupling to composition \cite{schmitt95}, and can exhibit an apparent
unstable constitutive relation, which would be mechanically unstable
under controlled strain rate conditions. Although there have been few
experiments on true lyotropic rigid rod systems in flow, wormlike
micelles can have a flow-induced nematic phase at higher
concentrations, and our results appear to qualitatively describe many
aspects of these experiments.  See, for example, the phase diagrams in
Ref.~\cite{CCD95}.

We have also shown schematically how our construction for coexistence
can be cast as an equivalent dynamical system, for which coexistence
corresponds to a heteroclinic saddle connection. In the most
general case stress selection depends on the nature of the gradient
terms in the dynamics \cite{jsplanar}, while in equilibrium systems the gradient terms
can be exactly integrated to yield a condition independent of the
gradient terms. The dynamical systems picture also yields an analogy
with a critical droplet which may prove promising in understanding the
non-equilibrium analogs of nucleation and growth.

{\bf Acknowledgments} We are grateful to M. Cates, B.~L. Hao, R. Ball,
and O. Radulescu for fruitful conversations.

\end{document}